# Can Hyperloops Substitute High Speed Rails in the Future?

Akhouri Amitanand Sinha[a] and Suchithra Rajendran[a,b,*]

[a]Department of Industrial and Manufacturing Systems Engineering, University of Missouri Columbia, MO 65211, USA
[b]Department of Marketing, University of Missouri Columbia, MO 65211, USA

Akhouri Amitanand Sinha
E-mail address: ask54@mail.missouri.edu
Telephone: 612-423-0394

*Suchithra Rajendran (Corresponding Author)
E-mail address: RajendranS@missouri.edu
Telephone: 573-882-7421

## Abstract

While the existing literature has focused on the impact of High Speed Rails (HSR) on the airline industry, we believe that this research is the first to examine the substitutability of HSR with Hyperloop services from an operational point of view. A simulation model is developed to compare the performance of both these alternate transportation modes for a network of three major cities in Europe (Amsterdam, Paris, and Frankfurt). Our results indicate that with a significantly lower pod capacity, the Hyperloop system will still be able to serve more customers compared to the HSR services, while the vehicle utilization is observed to be higher in the latter alternative for a given period of time. In addition, sensitivity analysis is conducted to assess the impact of variation in capsule capacity, number of pods in the system, and commuter variability. We further compare the two transportation modes with respect to their estimated infrastructure and operational costs as well as $CO_2$ emission. Finally, a cost-benefit analysis is conducted to estimate the passenger ticket price for Hyperloop services.

## 1. Introduction

In recent years, traffic congestion on highways is increasing, leading to economic loss, increased accidents, stress and pollution. A study by the American Transportation Research Institute (2018) concluded that traffic congestion on highways generated a loss in operational costs of over $74 billion to the trucking industries. Furthermore, they can experience an additional 250% on the transit cost per hour due to an unexpected congestion delay (US Department of Transportation, 2005). Another major impediment of expressway vehicle saturation is the disruption to the nation's supply chain (Hooper, 2018). This can impact the overall economy by incurring supplemental costs due to late deliveries while simultaneously requiring industries to maintain large inventories to compensate for the unpredictable nature of product arrival (US Department of Transportation, 2005). While it is apparent that improving the existing road facilities would be a good solution to address this issue, Winston and Langer (2006) observed that allocating funds to highways for constructing additional lanes would have minimal impact on the bottlenecks. Therefore, in recent years, with advancements in technology, several emerging transportation services, such as High-speed rails (HSR), Hyperloops, and Air Taxis, are evolving to improve the existing traffic condition. These facilities are expected to provide a faster and efficient mode of commute.

High speed rails, a type of mass transit, provide rapid energy-efficient transportation when compared to the traditional rail services (Zhou and Shen, 2011). Although these facilities were initially introduced in Japan, they are widely being used by major European countries, such as Germany, France and Spain. A typical HSR, has an average speed of approximately 150 mph on newer established tracks (Palacin et al., 2014). The seating capacity of a HSR vehicle is dependant on the design and ranges from over 400 passengers to approximately 1300 commuters per train (Givoni, 2006). Furthermore, it is observed that a HSR is nine times more energy efficient than airline facilities and nearly four times more efficient than driving on road (EESI, 2018).

While High-speed rails are being used by passengers in numerous European and Asian countries, Hyperloop services are emerging and are expected to operate in the forthcoming years. Hyperloop is based on the Maglev system, which utilizes magnetic properties to propel a pod towards its destination in a vacuum at high speeds. This eliminates the use of wheels in pods and thereby removing any friction from the track (Abdelrahman et al., 2017). In terms of operation, Hyperloops are comparable to a subway system but have fewer stops within a city. A typical Hyperloop system would have passenger capsules going in either direction inside two tubes of diameter 2.23 m. A pod would travel at a maximum speed of 750 mph and accommodate 28 passengers (Dudnikov, 2017; Rajendran and Harper, 2020). In contrast, a subway car has a capacity of approximately 54 seated commuters (Rajendran and Harper, 2020). Irrespective of capacity constraints, Hyperloop is expected to reduce the travel time between two cities drastically. For example, it is expected to cover the distance between LA and SF in about 35 minutes, which is faster than standard air transportation services. It would also have a boarding/departing time of 60 seconds, thereby retaining the fast turnaround time of a subway train. Figure 1 displays a typical design of a Hyperloop pod as proposed by SpaceX in 2013 (Musk, 2013).

Most previous literature has observed a positive impact of supplanting air transport with high-speed rails (HSR) (Castillo-Manzano et al., 2015; Takebayashi, 2014; Zhang et al., 2018). However, the present study focuses on investigating the substitutability of HSR with Hyperloop services. A discrete event simulation model is developed to compare the performance of both these alternate transportation modes for a network connecting the busiest airports and rail stations in Europe. For this purpose, we choose Paris (France), Amsterdam (Netherlands), and Frankfurt (Germany) (Eurostat, 2019). The juxtaposition between the Hyperloop and HSR is performed by considering several parameters, such as the number of pods (or rail cars), pod utilization (rail utilization), passenger cycle time, and overall lead time.

The remainder of the paper is structured as follows. The review of the literature pertaining to Hyperloop and high speed rails is presented in section 2. Methodology and model development are described in Section 3, while Section 4 details the data description. The results obtained for the base case along with sensitivity analysis are showcased in Section 5. Section 6 presents the discussion on comparing HSR with Hyperloops. Conclusion and future work are summarized in Section 7.

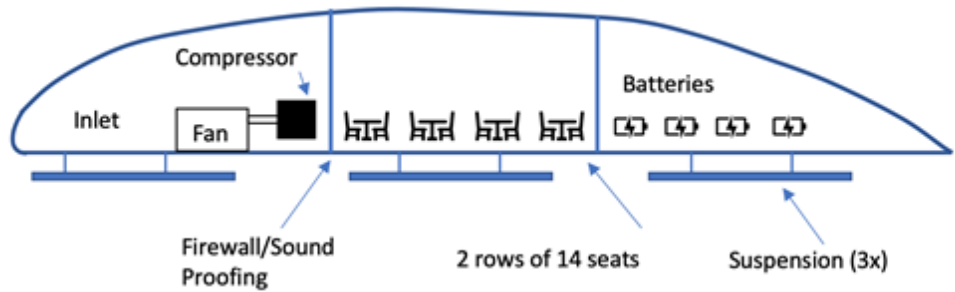

**Figure 1:** Typical Design of a Hyperloop Pod

## 2. Literature Review

This section reviews the work conducted in the literature pertaining to Hyperloop and HSR operations.

### 2.1 Hyperloop System

The Hyperloop system is expected to be a faster and economical alternative to conventional short-range aviation and high-speed rails. Moreover, a market study by NASA (Taylor et al., 2016) concluded that developing Hyperloop facilities would be cheaper than other high-speed railway networks. Recent research by Decker et al. (2017) focussed on investigating multidisciplinary characteristics affecting the vehicle design, such as thermodynamic, aerodynamic, electromagnetic, and energy analysis. The authors concluded that expanding the passenger capacity of pods from the original volume of 28 commuters would not have a significant effect on the overall cost. This would enable logistics companies to vary the pod length based on actual market demand. Bordone (2018) developed a framework for exploring socio-economic issues faced by Hyperloop in European Union using a qualitative study.

Several recent studies have focussed on the operational side of the hyperloop network. For instance, Rajendran and Harper (2020) developed a simulation model to analyze the impact of parameters, such as the total number of pods, pod capacity and demand variability, between San Francisco (SF) and Los Angeles (LA). Voltes-Dorta and Becker (2018) investigated the effect of establishing this service on airports in SF and LA using an exploratory analysis. They concluded that the Hyperloop system would provide a feasible alternative to air travel. Similarly, Santangelo and Andrea (2018) determined that while implementing the Hyperloop system is feasible, the initial costs of developing the necessary infrastructure are relatively higher than other modes of transportation.

**2.2 High-Speed Rails**

The emergence of high-speed rails has impacted the airline industry substantially on several aspects, such as environmental (D' Alfonso et al., 2018) and airfare and passenger demand (Chang and Lee, 2008; Suh et al., 2005). For instance, the introduction of HSR between the Frankfurt - Cologne route in 2002 led to an approximate 66% decline in the air passengers, eventually leading to the discontinuation of the air service (Clewlow et al., 2012). Similar trends were observed in other countries such as Japan (JR East, 2016), South Korea (Park and Ha, 2006), and China (Chen and Jiang, 2020; Zhang et al., 2017). Multiple researchers have focused on analyzing the complementarity and substitutability between the airline and high-speed rail industry (Castillo-Manzano et al., 2015; Sun et al., 2017; Wan et al., 2016; Zhang et al., 2018). However, studies are yet to consider the impact of emerging transportation services, such as Hyperloops.

Gundelfinger-Casar and Coto-Millan (2017) explored the implications of competition between air travel and HSR in Spain. They analyzed the demand for the two mediums as a function of commuter income, traveling price, and cost of alternative modes. They concluded that the latter two variables have a substantive impact on whether the HSR would substitute or complement airline services. Similarly, other factors, such as service frequency, travel time, and distance, were considered to be significant parameters affecting the viability of HSR over air transport in the literature (Behrens and Pels, 2012; Zhang et al., 2018). Chen (2017) utilized regression analysis for investigating the intermodal competition by integrating the supply and demand perspectives. The authors observed that high-speed rails had the maximum substitutional effect over air service between cities within the range of 500-800 km. Likewise, Gleave (2003) determined that HSR was not competitive for journeys less than 150-200 km and greater than 800-1000 km.

Yang and Zhang (2012) investigated the implications of competitiveness between air and high-speed rail transport based on profit, price, and social welfare. They observed a greater influence on public welfare by the HSR management resulted in a decrease in profit for both the transport mediums as they were competing for the same demand. They also concluded that the overall profits due to price discrimination between leisure and business travelers using HSR services remained unaltered. Whereas fewer business passengers utilize air transport due to cost differences when compared to leisure commuters. Adler et al. (2010) conducted a cost-benefit analysis for four trans-European networks using game theory. They found that upgrading the infrastructure for the entire European network would maximize social welfare and shift the demand from airlines to HSR.

Gonzalez-Savignat (2004) developed a simulation model to study the impact of HSR on the market share of the airline facilities. They observed travel time to be a significant variable in determining the market penetration, concluding that a rise in customer cycle time utilizing HSR would reduce the overall generated business. Similar findings were reported by Danapour et al. (2018), in which they created a discrete binomial logit model as a customer decision-making tool. The researchers noted fare price and trip duration to be the most influential parameters.

**2.3 Contributions to the Literature**

The contributions to the existing literature are multifold. While the existing literature has focussed on the impact of High Speed Rails (HSR) on the airline industry, we believe that this research is the first to examine the substitutability of HSR with Hyperloop services. In addition, we consider a multi-city Hyperloop network operation, whereas most previous literature is based on evaluating the system between two cities. Several parameters such as the number of vehicles in the system, passenger seating capacity, and commuter variability are varied to analyze the impact on vehicle utilization, passenger average cycle time, and lead time. We further compare the two transportation modes with respect to their estimated infrastructure and operational costs as well as $CO_2$ emission. Finally, a cost-benefit analysis is conducted to estimate the passenger ticket price for Hyperloop services.

## 3. Methodology

### 3.1 Model Description

Consider three cities ($\Phi_1$, $\Phi_2$, and $\Phi_3$) in the system with the customer $c$, commuting from one city to another by utilizing a travel mode represented by $m$, where $m \in \{h, r, f, c\}$.

1. $h$ – Hyperloop
2. $r$ – High-Speed Rail
3. $f$ – Airlines
4. $c$ – Conventional Rail

It is assumed that the passenger arrival rate for each hour $p$ by mode $m$, follows a Poisson distribution with a mean of $\lambda_{m,p}$. Further, the ticket price for Hyperloop is estimated to be greater than HSR and conventional rails but less than air transport, i.e., $t_f > t_h > t_r > t_c$. Let the total time to complete a trip via mode $m$ be depicted by $\theta_m^T$.

A customer $c$, would prefer traveling by $h$ if the following condition is met:

$$\mu_c \leq \rho_i$$

Where $\mu_c$ is a uniformly distributed random variable $U \sim (0, 1)$ and $\rho_h$ is the probability of the passenger switching from mode $i \in \{r, f, c\}$ to option $h$.

A customer would prefer substituting traveling via HSR and conventional rails with Hyperloops if they are willing to tradeoff the time saved ($\theta_h^T - \theta_i^T$), with the difference in the ticket prices ($t_h - t_i$) where $i \in \{r, c\}$. It is to be noted that certain air transport passengers might have multiple connecting flights and thus might not prefer switching to Hyperloop, and hence $\rho_f < 1$. Based on this information, the expected hourly demand for the Hyperloop service ($\lambda_{h,p}$) is given below.

$$\lambda_{h,p} = \sum_{i \in \{r,f,c\}} \rho_i \times \lambda_{i,p}$$

Now, since the sum of independent random Poisson variables is a random Poisson variable, it is expected that the passenger arrival rate per hour for the Hyperloop services would be Poisson distributed with mean $\lambda_{h,p}$.

## 3.2 Model Development

We develop simulation models to demonstrate the appropriateness of substitutability of HSR with Hyperloop services. The first step involves the customer's arrival in the system. Subsequent to their arrival at the station, they enter a queue and wait for the vehicle to pick them up. It is assumed that travelers do not balk (enter the facility and leave instantly) or renege (wait in the queue for a certain time period after entering the facility and then leaving without traveling) in both Hyperloop and HSR services.

To test the proposed models, we utilize the available flight and rail travel data between three major European cities - Paris, Amsterdam, and Frankfurt. The reason for particularly considering these three cities are as follows. According to Eurostat (2019), the Charles de Gaulle airport in Paris had over 76 million travelers in 2019. It was the maximum amongst all the airports in the European Union, followed by Schiphol airport in Amsterdam (72 million) and Main airport in Frankfurt (70 million). Similarly, Gare du Nord station in Paris is widely considered to be the busiest railway station in Europe (Dugdale, 2019), with approximately 214 million passengers availing of its facilities.

The performance of the Hyperloop and HSR are investigated based on comparing various parameters such as vehicle utilization, average commuter cycle time, and lead times. We define lead time as the difference between the traveler's actual and desired departure times (i.e., traveler lead time = actual departure time of customer at Station 1 – time that passenger desires to depart from Station 1). The cycle time is defined as the sum of the lead time and the customer travel time.

## 4. Data Description

Since Hyperloops are still in the design and testing phase, demand is estimated based on the procedure described in Section 3.1. In other words, the future market for the Hyperloop network is assessed based on the total number of commuters utilizing the existing flight and rail services between the three cities under consideration.

To anticipate the number of passengers traveling by air between Paris, Amsterdam, and Frankfurt, all flights serviced by major carriers was recorded. Flight details were obtained using the Google flights website (2019). The flight data from 2020 was not considered in this study due to the effects of the COVID-19 pandemic, which decreased customer demand (Budd et al., 2020). Subsequently, the total seating capacity information was recorded to estimate the maximum number of commuters flying from one city to another. Since the air travel demand is seasonal, this process is repeated for every hour for each day of the week to develop an accurate representation of the passenger demand. This establishes the upper limit on the number of travelers utilizing the air services. However, not every flight would be expected to function at 100% capacity. Therefore, the estimated upper limit was then multiplied by a parameter defined in the literature as the commuter variability (CVV <1) to generate an hourly rate table (Rajendran and Harper, 2020).

The process for determining the passenger demand by rail is similar to that of air travel. The number of trains commuting between the three cities was recorded based on their train schedules obtained from the EU Rail website (https://www.eurail.com/en/plan-your-trip/eurail-timetable). The maximum seating capacity based on the train type was then documented. It was observed that the trains operate at a far less frequently when compared to flights but have a much higher capacity. The travel times were noted, and the process identical to estimating customers transitioning from air travel mentioned above is repeated. Finally, the hourly rate table for each day of the week was created by multiplying it with the commuter variability parameter to obtain an estimated travel demand from each city. Table 1 depicts the average weekly passenger volume leveraging the air and rail facilities between the three cities. It was observed that a more significant percentage of commuters preferred traveling via the rail system when compared to the airlines. Furthermore, a larger deviation was noted for rail services, potentially due to less demand during the weekends. Figure 2 presents the average weekly demand experienced in various routes.

**Table 1:** Descriptive Statistics of the Air and Rail Traffic per Week

| City Pairs | Rail | | Air | |
|---|---|---|---|---|
| | Mean | Standard Deviation | Mean | Standard Deviation |
| **Frankfurt to Paris** | 1758 | 159 | 2164 | 205 |
| **Frankfurt to Amsterdam** | 2801 | 512 | 1700 | 103 |
| **Amsterdam to Paris** | 3370 | 563 | 1833 | 130 |
| **Amsterdam to Frankfurt** | 2540 | 537 | 1691 | 25 |
| **Paris to Frankfurt** | 1824 | 131 | 2117 | 277 |
| **Paris to Amsterdam** | 3195 | 734 | 1832 | 132 |

## 5. Results

### 5.1 Baseline Case

For the baseline setting, we set the seating capacity of Hyperloop to be 28 passengers in a single ride, whereas a single HSR vehicle can carry a maximum of 480 commuters, based on prior research. The commuter volume variability (CV) was assumed to be 0.8 from previous literature (Rajendran and Harper, 2020). Based on the grid search method, the number of Hyperloop pods in the system is set at 30. For the purpose of comparison, a similar number of rail cars was considered for the HSR services (even though the capacity of an HSR vehicle is substantially higher than that of a Hyperloop). The simulation models were executed for one week with 100 replications.

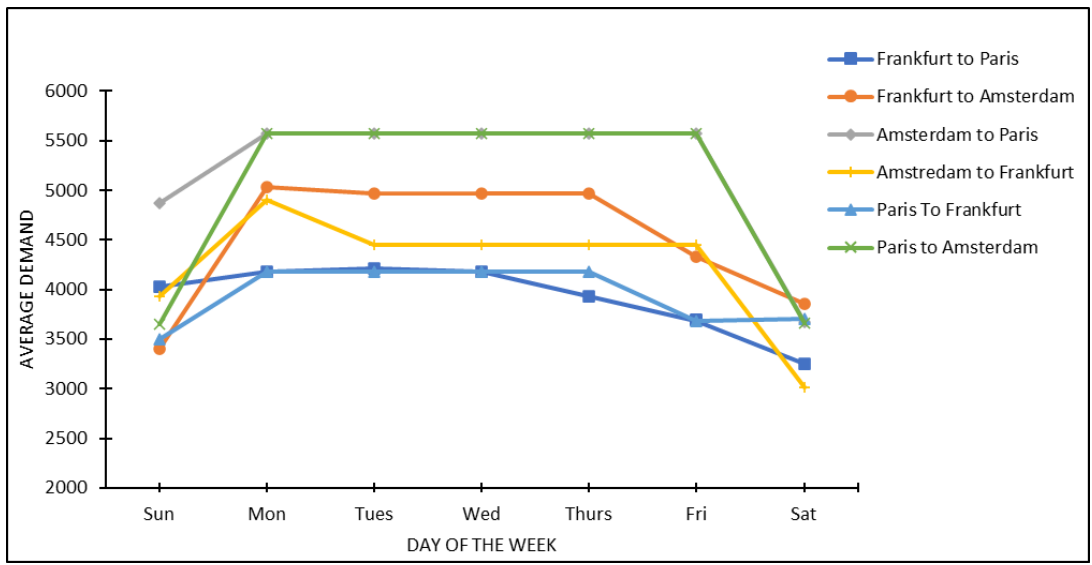

**Figure 2:** Average Demand Variation for Different Routes

The results for the baseline case comparing the performance of the Hyperloop and HSR services are showcased in Table 2. A Hyperloop pod travels at a higher speed and completes its journey substantially faster than any HSR. Therefore, a Hyperloop capsule is expected to remain idle for a longer time duration. This is supported by the findings in Table 2 as well, where it is observed that the vehicle utilization for HSR is greater than that of Hyperloop for all the routes. We can also note that the vehicle utilization for the journey between Amsterdam and Paris is approximately 10% higher when compared to other routes. This is because of the highest estimated passenger flow between these two cities, compared to the other pairs. Similarly, the commuters traveling between Amsterdam and Paris face over 50% and 70% greater cycle time and lead times, respectively, due to the highest passenger volume.

It is observed that the average HSR commuter's lead time is approximately 33% of the total cycle time. Moreover, the maximum capacity utilization is shown in the Frankfurt to Amsterdam route. This could be due to having the highest traveling time between any city that is considered in the present study, which would entail a greater vehicle utilization. However, the average customer lead time for this route is 40% less when compared to the maximum lead time in the system (which is noted for the Amsterdam to Frankfurt route). This could be primarily due to a more balanced commuter arrival distribution rate in the system. Furthermore, passengers going to Frankfurt face the highest cycle and lead times since the travel time from Paris and Amsterdam to Frankfurt is greatest compared to the other way around. Similarly, trains going to Amsterdam are on average 10% more utilized as they have a higher demand. In contrast, vehicles operating between Paris and Frankfurt are nearly 17% less used due to a low traveler demand, as described in Table 1. A paired sample t-test at 95% significance level showcases that all three performance metrics are significantly different.

Based on the results, it can be concluded that the Hyperloop system significantly outperforms HSR with respect to the performance metrics analyzed in the present study. Therefore, for a similar travel distance, HSRs could be substituted with Hyperloop services considering these measures. Nevertheless, there are other metrics, such as cost and sustainability, that have to be taken into consideration to evaluate the substitutability of HSR with Hyperloops. Therefore, we further compare the two transportation modes with respect to their estimated infrastructure and operational costs, as well as $CO_2$ emission, in Section 6.

**Table 2:** Comparison between Various Performance Metrics of Hyperloop and High-Speed Rails

| Travel Route | Vehicle Utilization (%) | | Cycle Time (minutes) | | Lead Time (minutes) | |
|---|---|---|---|---|---|---|
| | Hyp. | HSR | Hyp. | HSR | Hyp. | HSR |
| Frankfurt – Paris | 63.86 | 74.15* | 74.26 | 362.41* | 39.36 | 112.20* |
| Frankfurt – Amsterdam | 62.17 | 94.35* | 55.32 | 339.28* | 28.37 | 89.45* |
| Amsterdam – Paris | 72.33 | 89.12* | 115.03 | 308.06* | 125.10 | 110.62* |
| Amsterdam – Frankfurt | 59.76 | 90.47 | 54.08 | 399.84* | 27.13 | 152.54* |
| Paris – Frankfurt | 58.89 | 86.14* | 89.03 | 405.01* | 54.03 | 151.77* |
| Paris - Amsterdam | 67.76 | 93.57* | 165.14 | 320.80* | 135.60 | 121.68* |

Note: * denotes significance at 95% significance level

## 5.2 Sensitivity Analysis

In this section, various parameters, such as the number of vehicles in the system, commuter volume variability, and pod capacity, are varied and their impact on the performance metrics is examined.

### 5.2.1 Number of Hyperloop Pods in the System

As discussed earlier, the baseline setting has 30 Hyperloop pods in the system. In this section, the total number of capsules in the system (NIS) is varied linearly in steps of 10 from NIS#1 - NIS#4 (Table 3). Figure 3 (a) – (c) presents the results obtained for various settings. As expected, the overall vehicle utilization (Figure 3 (a)) decreased on average by approximately 40%, with a linear increase in the number of pods. While NIS#1 showcases a superior capacity utilization, the customer cycle time and lead time are observed to be extremely high between all three cities. Therefore, a rise in NIS led to a reduction in average commuter CT and LT by approximately 90%.

Passengers traveling between Amsterdam and Paris encounter the greatest CT (Figure 3 (b)) and LT (Figure 3 (c)) for NIS#1 and NIS#2. For NIS#1, the overall cycle time and lead time for the remaining routes are comparable with the results obtained by HSR services other than for those between Amsterdam and Paris.

**Table 3:** Sensitivity Analysis Settings

| Setting | Number of Pods in System (NIS) | Capsule Capacity (CC) | Commuter Volume Variability (CVV) |
|---|---|---|---|
| **Number of Pods in the System** | | | |
| NIS – 1 | 20 | 28 | 0.8 |
| NIS – 2 (Base) | 30 | 28 | 0.8 |
| NIS – 3 | 40 | 28 | 0.8 |
| NIS – 4 | 50 | 28 | 0.8 |
| **Capsule Capacity** | | | |
| CC – 1 (Base) | 30 | 28 | 0.8 |
| CC – 2 | 30 | 32 | 0.8 |
| CC – 3 | 30 | 36 | 0.8 |
| **Commuter Volume Variability** | | | |
| CVV – 1 | 30 | 28 | 0.7 |
| CVV – 2 (Base) | 30 | 28 | 0.8 |
| CVV – 3 | 30 | 28 | 0.9 |

On the other hand, commuting between Frankfurt and Paris displays over 25% higher CT and LT values for NIS#3 and NIS#4, which is counter-intuitive considering the fact that it has the lowest demand. An average decline of 15% in vehicle utilization can be noticed between NIS#2 and NIS#4. A two-sample t-test indicates a significant difference for all scenarios. Similarly, the average commuter cycle time and lead time were reduced by approximately 50% and 70%, respectively, between the two settings. Therefore, in order to regulate excess demand, management can decide to increase the number of pods, which will have a substantial impact on reducing the total travel time, thereby also improving the customer satisfaction rate.

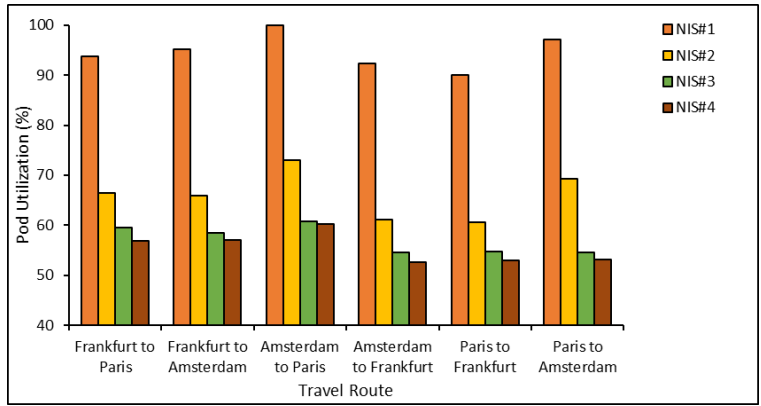

(a)

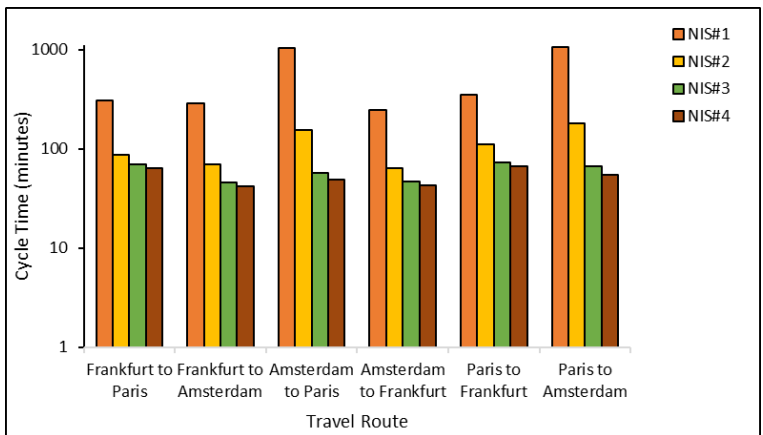

(b)

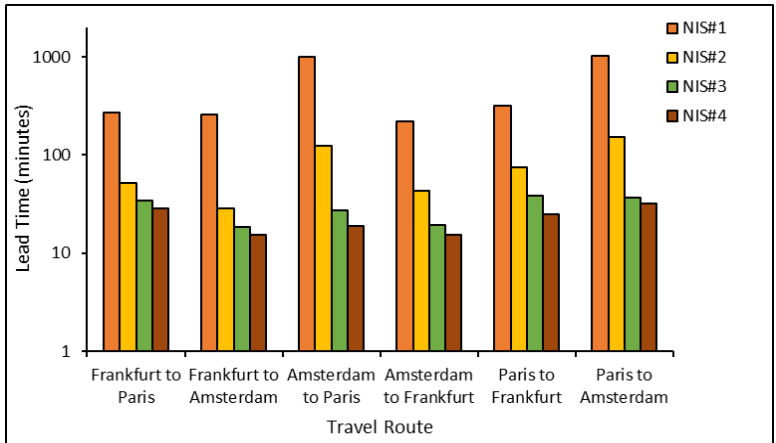

(c)

**Figure 3:** Impact of Variation of Number of Vehicles in the System on (a) Utilization, (b) Cycle Time, and (c) Lead Time

### 5.2.2 Hyperloop Capsule Capacity (CC)

In this section, the impact of increasing the capsule capacity (CC) on vehicle utilization (Figure 4 (a)), customer cycle time (Figure 4 (b)), lead time (Figure 4 (c)) is investigated. In the baseline case (CC#1), we assumed the pod capacity to be 28. Whereas each vehicle has a carrying capacity of 32 and 36 customers respectively for the subsequent cases, as shown in Table 3. As expected, the average vehicle utilization decreased by approximately 10% for all cities, with an increase in CC. A similar trend is identified for the customer cycle time and lead times with a decline of nearly 40% and 30%, respectively, from CC#1 to CC#3. The results obtained in the present study follow a comparable pattern as described by Rajendran and Harper (2020) for the Hyperloop operations between LA and SF.

The trip between Amsterdam and Paris showcases a maximum decline between CT and LT (~40%) due to an increase in CC from the baseline case (CC#1). A two-sample t-test establishes that the decline demonstrated by all parameters for CC#2 and CC#3 is significant at a 95% significance level. Similar to the baseline case (CC#1), Amsterdam to Paris has the highest capacity utilization, whereas Paris to Amsterdam has the maximum CT and LT for CC#2 and CC#3. Based on the results from two-sample t-tests, it can be concluded that the rate of change in the hyperloop capsule capacity has a significant impact on the performance metrics.

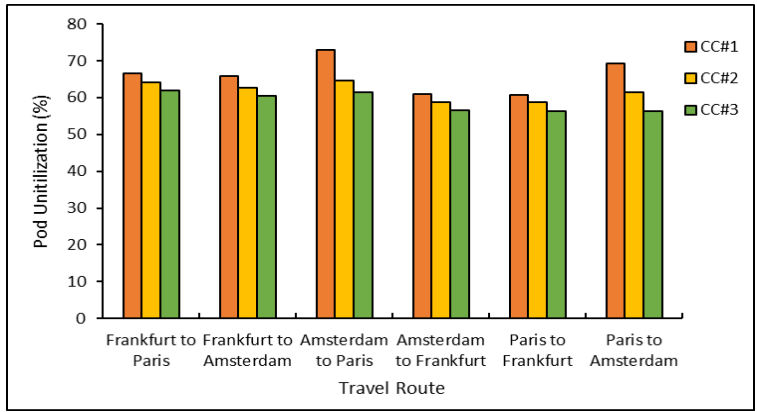
(a)

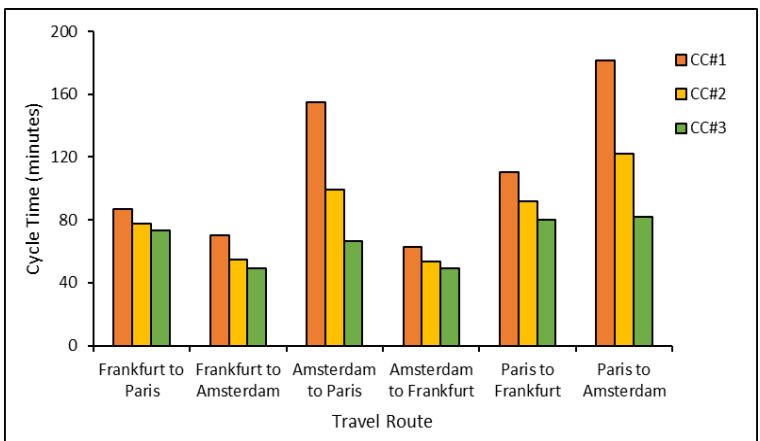
(b)

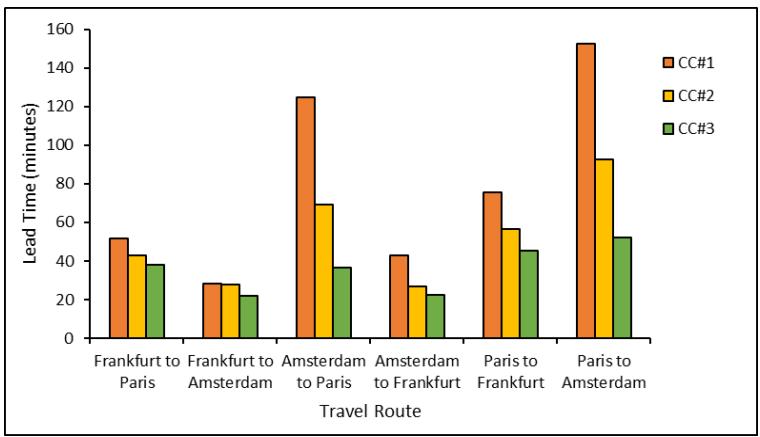
(c)

**Figure 4:** Impact of Variation of Capsule Capacity on (a) Utilization, (b) Cycle Time, and (c) Lead Time

### 5.2.3 Commuter Volume Variability (CVV)

It is expected that not all eligible passengers would be willing to avail of the Hyperloop services. Therefore, the commuter volume variability (CVV) is modified to capture the effect of altering overall demand on the performance parameters. While the baseline case (CVV#2) is set at 80%, similar to Rajendran and Harper (2020), the other scenarios are modified in steps of 10% with CVV#1 and CVV#3 evaluating the demand at 70% and 90% respectively (Table 3). Figure 5 (a) displays the impact of changing CVV on pod utilization. Similarly, Figures 5 (b) and (c) depict the effect of varying the parameter on CT and LT, respectively.

It is observed that a decrease in CVV led to a reduction in cycle time for customers traveling the Amsterdam - Paris route by over 50 minutes. Whereas an approximately 30% increase in CT is observed for the same journey in CVV#3. A similar trend is noticed for the lead time as well. Even though transitioning from CVV#1 to CVV#3 increments vehicle utilization by 10% on average, the CT and LT rise close to 40% and 50%, respectively, showcasing that CVV is an important factor that affects the system efficiency and client satisfaction. Similar to our previous analysis, a two-sample t-test indicates that the results presented in Figure 5 (a) – (c) are significantly different from each other at a 95% significance level.

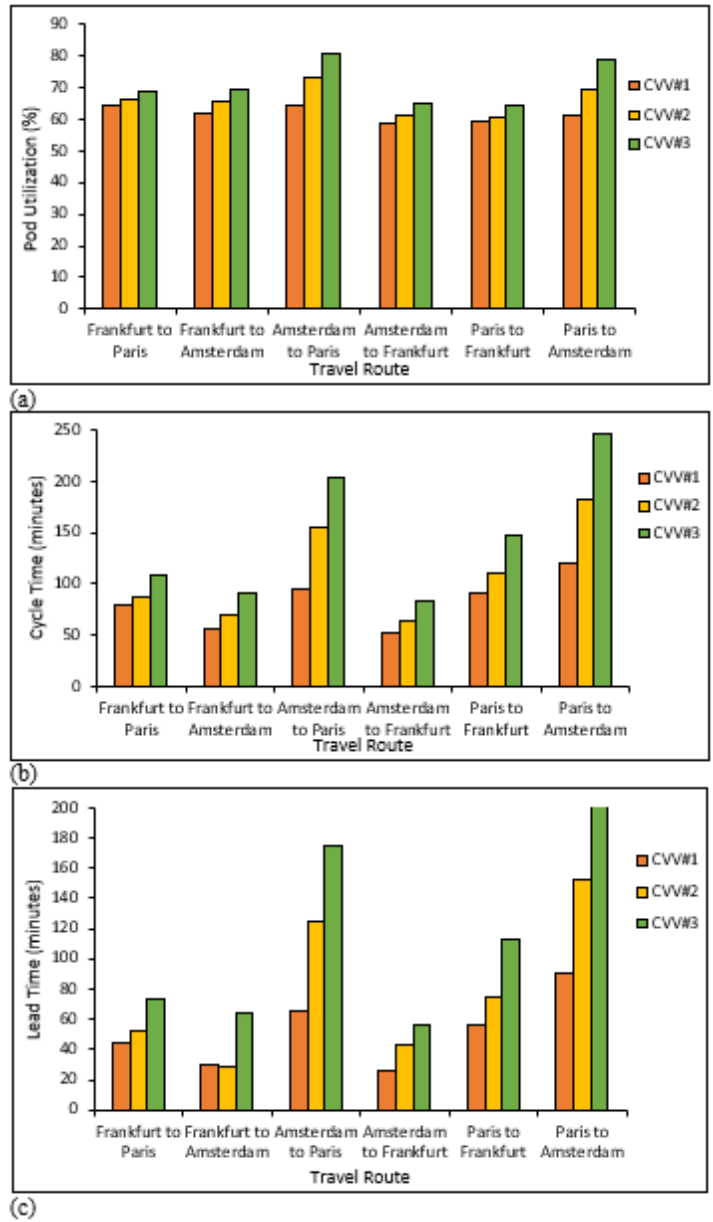

**Figure 3:** Impact of Variation of Commuter Variability Parameter on (a) Utilization, (b) Cycle Time, and (c) Lead Time

## 6. Discussion

The results obtained in the previous section showcase the efficiency of the Hyperloop services over HSR based on time savings and vehicle utilization. However, the feasibility of substituting HSR with Hyperloops is also contingent on several other critical parameters, such as infrastructure and operational costs, sustainability, and ticket prices. These factors are discussed in detail in this section.

### 6.1 Infrastructure Costs

Similar to other emerging transportation services, several cost components such as infrastructure and operational costs are involved in the Hyperloop system. These costs would have a significant influence on the overall travel fare, which could, in turn, lead to a decline in the commuter's willingness to utilize this emerging service. The infrastructure costs include the expenses associated with developing new stations, constructing tunnels between each city, land acquisition, and environmental planning (Taylor et al., 2016). Musk (2013) estimated the initial capital costs for the Hyperloop technology to be approximately $17 million per mile for the LA-SF route. On the other hand, the infrastructure development costs for the high-speed rail network for the same route are evaluated at $65 million per mile (Taylor, 2016). Nevertheless, the existing literature indicated that the actual infrastructure costs for both the services would be more analogous to each other due to the Hyperloop system requiring expenditure for constructing tube and vacuum pumps (Hansen, 2020; van Goeverden et al., 2018).

## 6.2 Operational Costs

The operational costs comprise the employee costs, maintenance and system control, etc. The California High-Speed Rail Authority (CHSRA) estimated the maintenance cost for the LA-SF corridor to be approximately $200,000 per mile and incurring $83.22 per revenue service hour (CHSRA, 2012). While the literature on similar parameters for the Hyperloop facility is not available, a previous study by van Goeverden et al. (2018) suggested pod capacity to be a potential limitation for the Hyperloop system affecting its financial performance. Furthermore, Musk (2013) suggested an electronic ticketing system to reduce staffing costs. However, further research integrating various parameters, such as utility and IT costs, customer service, security, etc., would be required before juxtaposing it with the HSR network.

## 6.3 Sustainability

It is expected that the Hyperloop services would be energy efficient by more than two times when compared to HSR due to low track friction, less air resistance, and utilization of solar panels (Taylor, 2016). A recent study by Janic (2020) observed comparable $CO_2$ emissions (approximately 40 gCO2/s-km) between HSR and Hyperloop. They also concluded that the $CO_2$ emission level was substantially lower than a standard aircraft (100 gCO2/s-km). The energy consumption and $CO_2$ emissions by both services decrease proportionally with increasing travel distance and seating capacity (Janic, 2020). Similar research by Decker et al. (2017) observed that the energy consumption costs due to increasing the seating capacity remained relatively constant. Therefore, the decision-makers can consider expanding the capaciousness of Hyperloop pods for maximizing demand fulfillment.

## 6.4 Cost-Benefit Analysis

It is expected that the travel cost incurred by the Hyperloop passengers would be relatively higher during the initial stages of operation. In this section, we perform a cost-benefit analysis (CBA) for an ex-ante evaluation between the two transportation modes to suggest the optimal pricing structure to attract passengers to utilize the Hyperloop facility over HSR.

The total cost $tc_m$ incurred by the customer traveling from mode $m$ for any pair of cities is given by Equation (1).

$$tc_m = t_m + (V \times \theta_m^T) \qquad (1)$$

Where, $C_i$ is the average ticket cost in transportation alternative $m$ between two cities in dollars, $V$ is the value of time for passengers in dollars/hour and $\theta_m^T$ is the total travel time between the two cities using alternative $m$ in hours. Previous literature suggests that the time value is a function of the wage rate (Hultkrantz, 2013; Wardman et al., 2016; Zhao et al., 2015).

The total cost for HSR and Hyperloop passengers are given by Equations (2) and (3), respectively.

$$tc_h = t_h + (V \times \theta_h^T) \qquad (2)$$
$$tc_r = t_r + (V \times \theta_r^T) \qquad (3)$$

Equation (4) needs to be satisfied in order for Hyperloop services to be comparable with HSR in terms of costs experienced by the customers.

$$tc_h \leq tc_r \qquad (4)$$

Substituting Equations (2) and (3) in Equation (4) we can analyze the ideal ticket price for availing the Hyperloop facilities, as shown by Equation (5).

$$t_h \leq t_r + (V \times (\theta_r^T - \theta_h^T)) \tag{5}$$

The total travel time for each pair of cities (i.e., difference between cycle time and lead time) is calculated from the results shown in Table 2. The ticket cost of high-speed rails is estimated based on the seven-day average prices available on the EU rail website. The time value is obtained from the meta-analysis conducted by Wardman et al. (2016).

The price difference between the two transportation modes is shown in Table 4. It is observed that the travel fare for Hyperloop services exceeds the current ticket cost for HSR services by approximately 78%. Furthermore, the price suggested in the present study for Hyperloop services is comparable to the existing airline operations while also being considerably higher than the originally proposed price structure of USD 20 (Musk, 2013). Therefore, in order for the emerging technology to become comparable in terms of commercial and marketability values, a significant reduction in the ticket costs would be necessary for the future.

**Table 4:** Ticket price difference between the Hyperloop and HSR services

| City Pairs | C HSR ($) | Vt ($/hr.) | C HL ($) |
|---|---|---|---|
| Frankfurt to Paris | 18.86 | 35.49 | 146.12 |
| Frankfurt to Amsterdam | 49.00 | 35.49 | 173.21 |
| Amsterdam to Paris | 21.55 | 40.55 | 138.33 |
| Amsterdam to Frankfurt | 56.50 | 40.55 | 198.42 |
| Paris to Amsterdam | 38.96 | 32.04 | 131.34 |
| Paris to Frankfurt | 16.00 | 32.04 | 130.81 |

The following managerial insights are proposed based on the results obtained in the present study:

i. A linear increase in the number of pods in the system leads to an exponential decline in vehicle utilization, passenger cycle time, and lead time. Therefore, it is recommended to determine the optimal number of pods in the system for maximum customer satisfaction.

ii. All parameters showcase a linear decrease with a linear increase in the seating capacity of the pod. Based on a previous study by Decker et al. (2017), capsule capacity can be increased without a significant impact on the overall structural cost. Therefore, it is suggested to perform a cost-benefit analysis of expanding the number of seats in the pod. Consequently, if the logistic companies decide to decrease the number of vehicles in the system, then the supply shortfall could be offset by increasing the capsule capacity.

iii. It is observed that a linear increase in demand results in all three performance metrics rising linearly. Thus, when the latent demand arises, it is proposed to immediately increase the number of capsules in the system. This would mitigate the customer density growth for the facility by reducing the cycle time and average commuter lead time.

iv. The average lead time for passengers availing of the high-speed rail facilities is significantly higher than the Hyperloop services. Consequently, introducing additional trains would aid in reducing the long lead times.

v. In order to increase the customer willingness to utilize the Hyperloop services, it is recommended to reduce the ticket price by approximately 80% in the future based on the results obtained from the cost-benefit analysis.

## 7. Conclusion and Future Work

Traffic congestion has led to a significant loss of productivity and increased the cost of travel. Several researchers and practitioners are examining emerging transportation methods that are economically viable and significantly reduce transit time, such as Hyperloops, air taxis, and high-speed rails (HSR). Previous literature has concluded that HSR have a considerable impact on traditional airline services over short and medium distances with respect to cost and ride time. To the best of our knowledge, this study is the first to focus on exploring the substitutability of high-speed rails with Hyperloop systems, which are expected to commence services in forthcoming years for equivalent distances. Simulation models are developed in the present work to compare the overall customer time in system, lead time, and vehicle utilization between Hyperloops and HSR between three major European cities. Furthermore, we juxtapose the two transportation means based on several categories such as sustainability, infrastructure and operational costs. We also estimate the average ticket prices for commuters utilizing the Hyperloop services through a cost-benefit analysis.

The base case results showed that passengers would experience approximately 75% and 34% decrease in cycle time and lead time while commuting through the Hyperloop system compared to HSR services. However, pod utilization is nearly 25% lower than the HSR services due to its higher speed. It is observed that Hyperloop customers transporting between Paris and Amsterdam would encounter the greatest CT and LT. On the other hand, HSR users traveling to Frankfurt experience the highest performance metrics when compared to other routes. Sensitivity analysis is performed to investigate the impact of various alternate Hyperloop scenarios, such as a change in the number of pods in the system, capsule capacity, and commuter variability parameter. The outcomes indicate a significant influence of the three parameters on all the performance measures.

The proposed simulation model can serve as a decision support tool for any logistic companies interested in advancing into the Hyperloop business. Typically, a commuter is expected to compare the feasibility of availing a service based on various factors such as price, distance, perceived safety, and familiarity. Therefore, a major limitation of this study is that it does not consider the impact of customer willingness to utilize the Hyperloop services based on these factors. Future work could investigate the influence of these parameters on demand variation. Similarly, the effect of multiple socio-economic criteria affecting usability is not included in the current research. Therefore, multiple criteria models can be developed in the future to generate tradeoff alternatives. While the present study proposes reducing the ticket price significantly to make the Hyperloop facility more comparable with HSR, future work can conduct a more detailed investigation by considering the impact of infrastructure, overhead, and maintenance costs. Another major drawback of the current research is that it does not consider the impact of introducing Hyperloops on the existing air and rail services. Thus, future work can explore the differences between cooperation and competition between these services on the market.